\documentclass[12pt, preprint]{aastex}

\shorttitle{Discovery of hot gas in outflow in NGC3379}
\shortauthors{Trinchieri, Pellegrini,Fabbiano et al.}

\def\ergs{erg s$^{-1}$}

\def\Reff{$R_{\rm e}$}

\begin{document}


\title{Discovery of hot gas in outflow in NGC3379}


\author{G. Trinchieri}
\affil{INAF-Osservatorio Astronomico di Brera, Via Brera 28, 20212 Milano, Italy}
\email{ginevra.trinchieri@brera.inaf.it}

\author{S. Pellegrini}
\affil{Dipartimento di Astronomia, Universita’ di Bologna, Via Ranzani 1, 40127 Bologna, Italy}

\author{G. Fabbiano, R. Fu, N. J. Brassington, A. Zezas, D-W. Kim}
\affil{Harvard-Smithsonian Center for Astrophysics, 60 Garden St., Cambridge, MA 02138}

\author{J. Gallagher}
\affil{Department of Astronomy, University of Wisconsin, Madison, WI 53706-1582}

\author{L. Angelini}
\affil{Laboratory for X-ray Astrophysics, NASA Goddard Space Flight Center, Greenbelt, MD 20771}

\author{R. L. Davies}
\affil{Sub-Department of Astrophysics, University of Oxford, Oxford OX1 3RH, UK}

\author{V. Kalogera}
\affil{Northwestern University, Department of Physics and Astronomy, Evanston, IL 60208}

\author{ A. R. King}
\affil{Theoretical Astrophysics Group, University of Leicester, Leicester LE1 7RH, UK}

\and

\author{S. Zepf}
\affil{Department of Physics and Astronomy, Michigan State University, East Lansing, MI 48824-2320}



\begin{abstract}
We report the discovery of a faint ($\rm L_x \sim 4 \pm 1.5 \times 10^{37}$
\ergs, 0.5-2 keV), out-flowing gaseous hot interstellar medium (ISM)
in NGC 3379.  This represents the lowest
X-ray luminosity ever measured from a hot phase of the ISM in a nearby 
early type galaxy.  The discovery of the hot ISM in a very deep {\it Chandra}
observation was possible thanks to its  unique spectral and
spatial signatures, which  distinguish it from the integrated stellar
X-ray emission, responsible for most of the unresolved emission in the
{\it Chandra} data.  This hot component is found in a region of $\sim
800$ pc in radius at the center of the galaxy and has a total mass 
M$\sim 3 \pm 1 \times 10^5$ M$_{\sun}$.  Independent theoretical prediction
of the characteristics of an ISM in this galaxy, based on the intrinsic properties 
of NGC~3379, reproduce well the observed luminosity, temperature, and radial
distribution and mass of the hot gas, and
indicate that the gas is in an outflowing phase, predicted by models
but not observed in any system so far. 
\end{abstract}





\section{Introduction}

The nature of the X-ray emission of elliptical galaxies has been
debated since their discovery as X-ray sources with the {\it Einstein
Observatory}, which led to the early suggestion of ubiquitous hot X-ray
emitting halos in hydrostatic equilibrium  (Forman et al. 1985). These halos would
originate from the stellar ejecta resulting from their normal evolution,
accumulated during the lifetime of the galaxies, and
retained by their deep dark matter potentials. However, from the analysis
of these early observations and of ROSAT and ASCA data, a more complex
picture of the X-ray emission emerged, suggesting that the
non-nuclear X-ray emission of these galaxies consist of varying amounts
of hot ISM, and of a baseline component related to the stellar
population, either from the finale stages of stellar evolution (i.e. 
low-mass X-ray binaries, LMXBs, 
Trinchieri \& Fabbiano 1985; Canizares et al. 1987; Kim, Fabbiano \&
Trinchieri 1992; Fabbiano et al. 2006; Matsushita et al. 1994) or possibly
from coronal stellar emission 
(Pellegrini \& Fabbiano 1994). Although controversial at
the time, this picture has important implications for the evolutions
of the stars and of the gaseous components of ellipticals, suggesting
halo retention in the most massive systems, and partial or full winds
in most galaxies (Ciotti et al, 1991; David et al. 1991; Pellegrini \& Ciotti 1998).

With {\it Chandra} for the first time the X-ray emission of nearby
ellipticals can be imaged, and the existence of the different emission
components can be pursued by direct observational means. These
observations have led to the detection and study of LMXB populations
(see review in Fabbiano 2006), and the removal of their contamination
from the diffuse emission component (e.g. NGC~1316, Kim \& Fabbiano 2003).
In some galaxies this diffuse emission is due
to gaseous halos; however, in other galaxies, where the
X-ray luminosity function can be probed to very low luminosities and the residual
emission is itself of low luminosity, unresolved LMXBs
and possibly stellar emission may account for a large amount or even the
bulk of the X-ray luminosity (e.g.NGC821, Pellegrini et al 2007a,b; M32,
Revnintsev et al. 2007).

We can now attempt to constrain  the presence and properties of vestigial
hot halos in these gas-poor galaxies. Since ellipticals are
expected to host super-massive nuclear black holes (e.g., Richstone et al. 1998), which are
however radiatively quiescent in most cases (Fabian \& Canizares 1988, Pellegrini 2005, Ho 2008),
these measurements provide useful constraints
for models of halo evolution and nuclear feedback (e.g., Springel, Di Matteo \& Hernquist
2005; see Pellegrini et al 2007b).  The nearby, gas-poor, early-type galaxy
NGC3379 provides an excellent target for these investigations. 
This is a very well studied prototypical elliptical, with $\rm L_B = 1.5 \times 10^{10}$
$\rm L_{B \sun}$. A
series of very deep  exposures were obtained for this galaxy by our team
(PI: Fabbiano), as part of a very large {\it Chandra} program aimed
at unraveling and studying the different X-ray emission components of
ellipticals. Together with  pre-existent archival data (David et al
2005),
these observations resulted in an integrated exposure of 337 ks with ACIS
S-3 (see Brassington et al. 2008 for details).
In this paper we present our analysis of the diffuse/unresolved
X-ray emission component of this galaxy. This analysis leads us to
conclude that in addition to the unresolved emission of faint LMXBs and
other stellar sources (Revnivtsev et al. 2008), a hot ISM is likely to
be present in the central 800 pc of NGC3379.

We discuss our data analysis in Section 2, and our interpretation of
these results in Section 3. Our conclusions are summarized in Section 4.
We assume a distance of 10.6 Mpc (Tonry et al. 2001), which gives a
scale of 51 pc/\arcsec.

\section{Analysis and Results}

We have used the same dataset that Brassington et al. (2008) 
used to determine the
source list and compose the most complete X-ray catalog to-date in NGC
3379, down to  (0.3-8.0) keV luminosities $\rm L_x \sim 10^{36}$ \ergs.
All sources in the catalog
have been excluded from the present analysis.  We have
used an aperture of 2$''$ radius.
On axis, this aperture encircles $>$ 97/95\% of the Point Spread Function (PSF), for energies 0.5/1.5
keV (see Figures  4.6 and 4.20, and Table 4.2 in The Chandra Proposers' Observatory
Guide, and our own simulations with ChaRT, see Appendix), and it degrades very slowly to encircle $\sim 90$\% at $\ge 3'$ for the same energies (Fig. 4.13 in the The Chandra Proposers' Observatory
Guide).   As will become apparent in what follows,
this work  will concentrate on the central area NGC~3379, of $\sim 80''$
radius; therefore, our data are not affected by the
degradation of the  {\it Chandra} point response function, and we do not need to consider a larger area for source exclusion. We discuss the possible contamination of the PSF in the Appendix.

The size of the area considered also ensures a very homogeneous coverage
from all five observations, since the region under study resides
well within the  area uniformly covered in all the {\it Chandra}
exposures (roughly a $8' \times 5'$ rectangle, see Fig. 1 in
Brassington et al. 2008).  Therefore we do not need to resort to modelling
of the exposure, and we can use a local estimate of the background from
the same merged dataset.  This procedure minimizes the uncertainties introduced
by using a template or ``blank sky observations" of different portions
of the sky at different times, to estimate the field background.

We have used two different merged datasets for our analysis. 
All five observations were
used for the spatial analysis 
but only the four most
recent observations were merged for the spectral analysis, given the
significantly different response of the  ACIS S3 CCD in the first observation
obtained in 2001 (AO2).  Since this  observation has the shortest
exposure, not using it will not significantly degrade the statistics,
while reducing the systematic error from  the calibration uncertainties.
The responses of the last four observations over the 
area of interest are virtually identical

\subsection{Radial Distributions of the X-ray Surface Brightness}

\begin{figure}
\plottwo{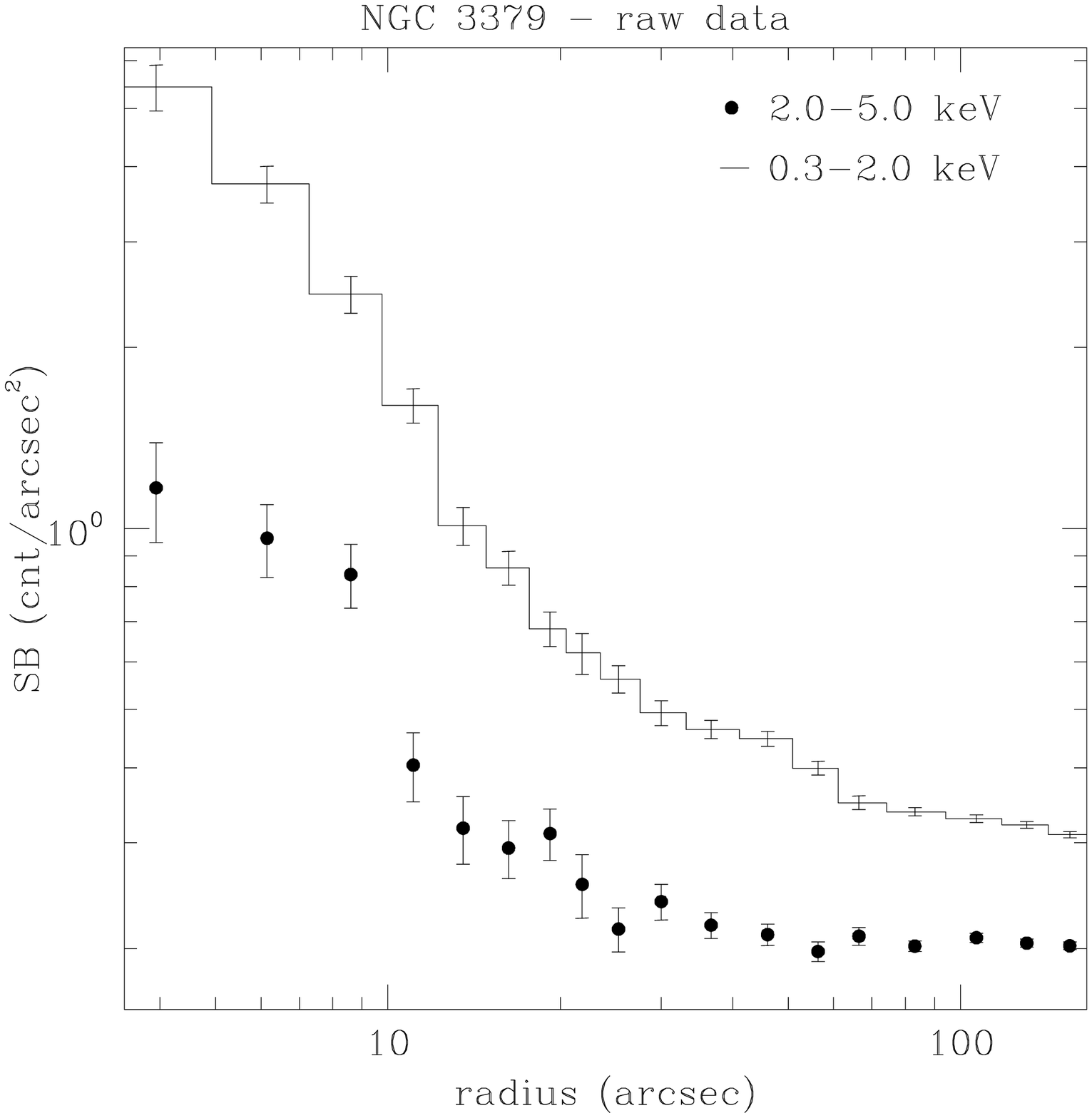}{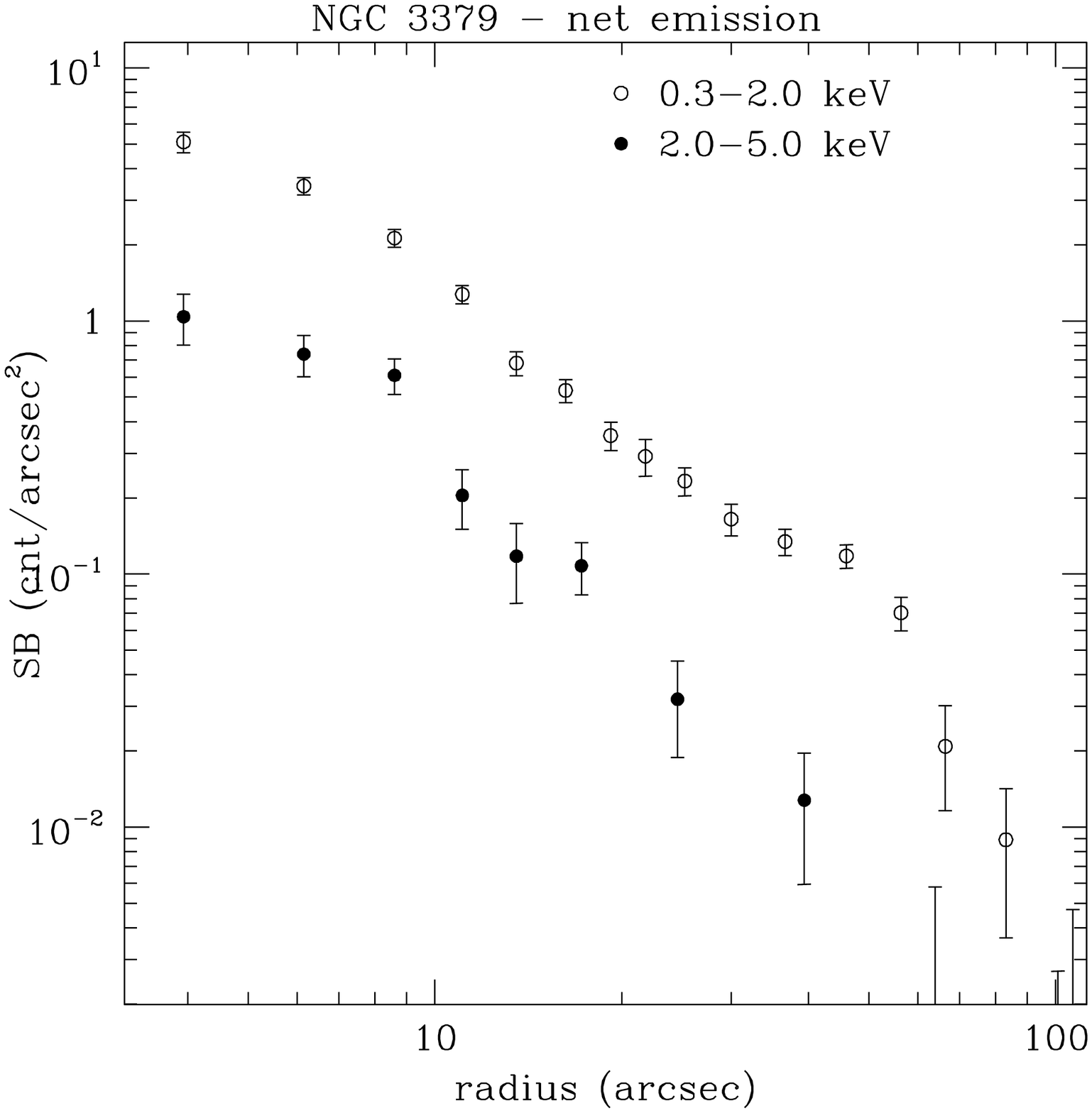}
\caption{Azimuthally averaged radial profiles of the unresolved
emission in NGC 3379, in two broad energy bands.  Left: raw data; right:
background subtracted data}
\label{raw}
\end{figure}

Fig.~\ref{raw}  shows azimuthally averaged  radial profiles of the
diffuse X-ray emission in the  (0.3-2.0 keV) and (2.0-5.0)~keV
energy bands. The choice of the spectral ranges is  motivated
by our intention to discriminate between a harder band (2.0-5.0)~keV,
where all the emission, resolved and unresolved,  is likely to arise from
stellar sources, and a softer band (0.3-2.0)~keV, to which a hot ISM may
contribute. We have excluded all detected sources (as explained
above), and centered the concentric annuli
on the nucleus, identified as the optical/IR
center and coinciding with source \# 81 in
Brassington et al. (2008).  Note that as a consequence of this, and also 
due to source crowding at the center, the X-ray profiles cannot probe the very central region
within r=2--3\arcsec. The raw 
profiles shown in Fig.~\ref{raw}  flatten at a radius
of $> 80''$, indicating  that the field background dominates.   We can
therefore use the 100$''-140''$ region to estimate the local background
and subtract it from the emission, to produce the net profiles shown in
Fig.~\ref{raw}-right in the
same energy bands. Note that the soft profile is more extended and 
more centrally peaked 
relative to the hard profile. 

\begin{figure}
\plotone{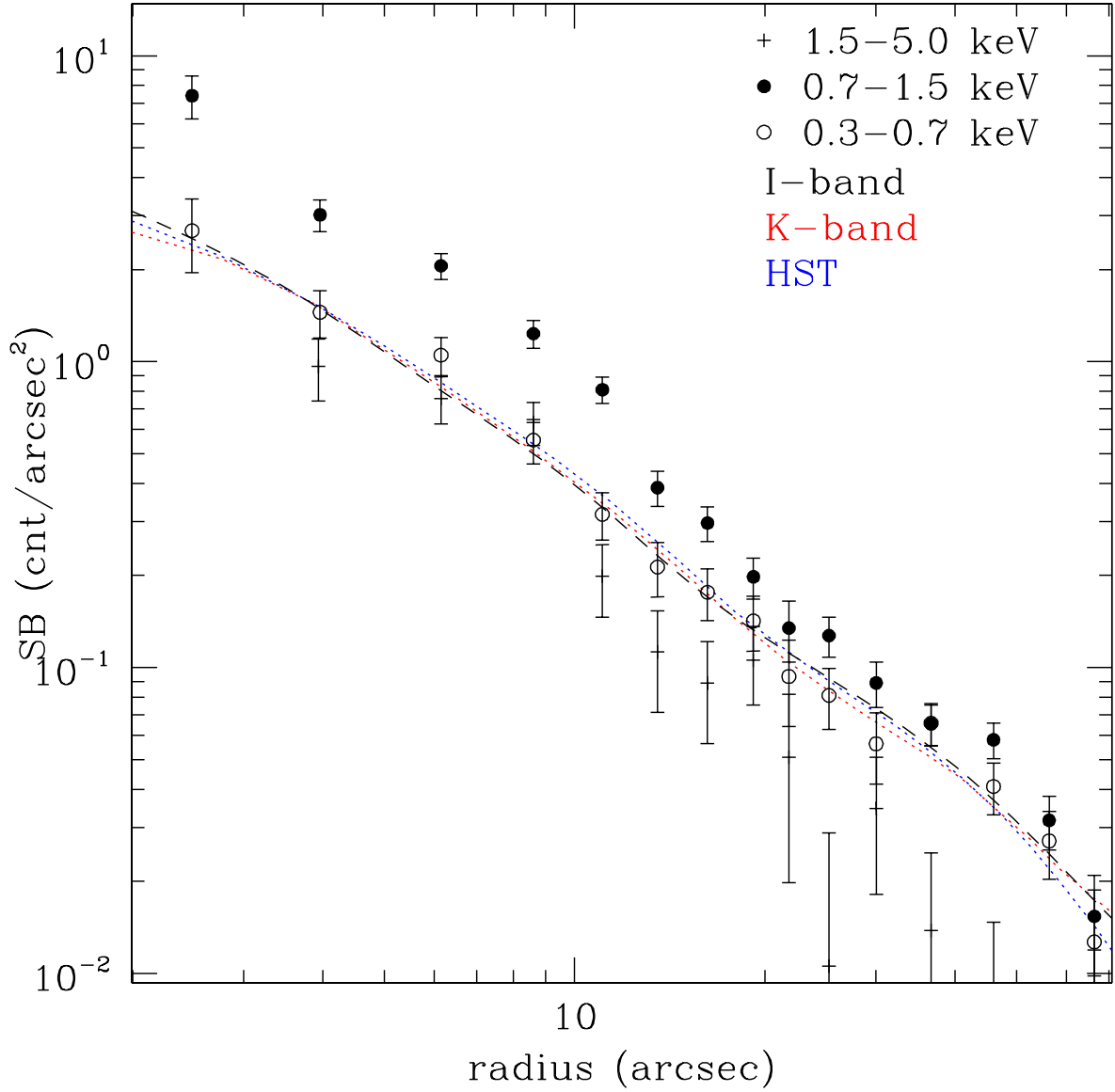}
\caption{Azimuthally averaged net profile in different X-ray bands,
compared to the optical and K-band profiles}
\label{softprof}
\end{figure}

In Fig~\ref{softprof} we divide the full energy range considered in
three, and show the azimuthally averaged net profiles in
the (0.3-0.7) keV, (0.7-1.5) keV and (1.5-5.0) keV bands.  This choice
of  energy boundaries  was motivated by the results of the spectral 
analysis of the diffuse emission (see next section, Fig~\ref{unfold}),
to maximize the contribution of the two optically thin thermal plasma
component identified in the X-ray spectrum, with temperatures of  $\sim
0.3$ keV and  at  $\sim 1$ keV, respectively.  It is clear from Fig.~\ref{softprof} 
that while both the very soft and the harder component follow a similar radial
distribution, the soft (0.7-1.5 keV) component has a distinct excess for galactocentric 
radii r$\le 20''$.   

In the same Fig.~\ref{softprof} we plot the radial profiles in the
optical and in the near IR K-band, from the HST F814 filter image
and the  2MASS k-band image, obtained from their respective archives
(see also Cappellari et al. 2006).
These profiles are also azimuthally averaged, but do not cover the full
360\degr\ azimuthal plane of the galaxy, except for the very center
($10''$ and $30''$ radius for HST and 2MASS respectively) since the two
images only partially cover the galaxy. 
We also add the Sauron I-band
data profile from Cappellari et al. (2006) that cover instead the full
extent of the galaxy. The profiles are normalized to the X-ray data by
rescaling them for radii r$>100''$.

With the exception of the very central region ($<3''$ radius), the
optical-IR profiles follow one another, and trace well the distribution
of the very soft and hard X-ray profiles.  On the contrary, the (0.7-1.5) keV
profile is more centrally peaked and only becomes consistent with the shape
of the optical-IR ones at r$>15''$.  

\subsection{Spectral Analysis}

We have analyzed the spectral data extracted in the  $2''-15''$,
$2''-30''$ and  $2''-45''$ regions from the merged dataset from the four more
recent observations.  Most of the results discussed here come from the
$2''-30''$ region, which is a reasonable compromise that includes a large
fraction of the diffuse emission at high significance, but all regions 
studied give a consistent picture.  To increase the
statistical significance of each spectral bin, we have binned the data
to obtain a minimum significance of 2$\sigma$ or better in the net data.
The spectral fits are done assuming the line-of-sight galactic N$_H$
of 2.7$\times 10^{20}$ cm$^{-2}$, the abundance tables of Wilms et al.
(2000),  the APEC model to account from  the optically thin thermal
emission  of a plasma, with abundances fixed at the 100\% value (different
values do not change the final results), and a
Bremsstrahlung to account for the X-ray emission of unresolved fainter
LMXBs and stellar X-ray sources.

\begin{figure}
\plotone{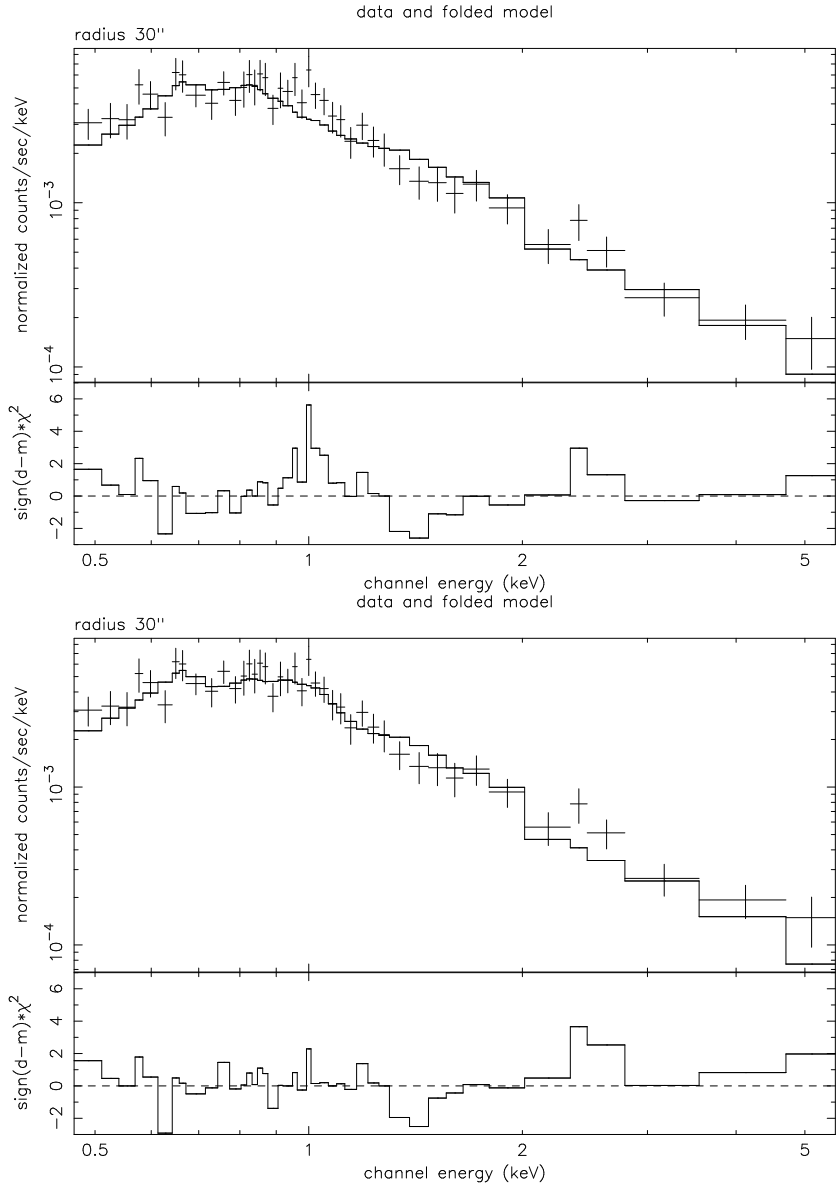}
\caption{Spectral data in the 30\arcsec\ region, fitted with a single
temperature (0.3 keV) and a two temperature APEC model (top and bottom 
respectively, see text),
plus a 7 keV bremsstrahlung. For displaying purposes, data are rebinned to 5$\sigma$. 
}
\label{spec1}
\end{figure}

\begin{figure}
\plotone{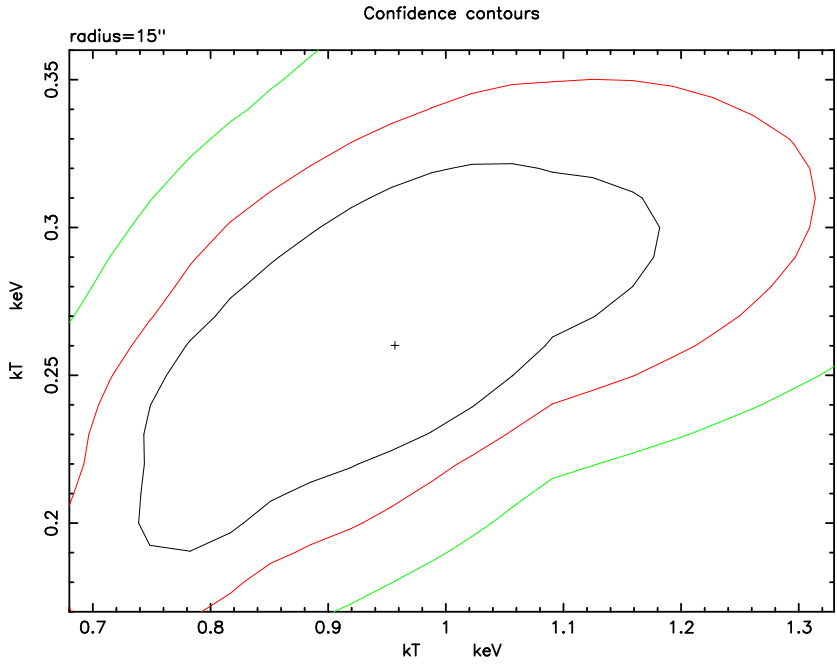}
\caption{Confidence contour regions
(at 68, 90 99\% level for two interesting parameters)
for the two temperatures of the APEC component (see text).
}
\label{contour}
\end{figure}

\begin{figure}
\plotone{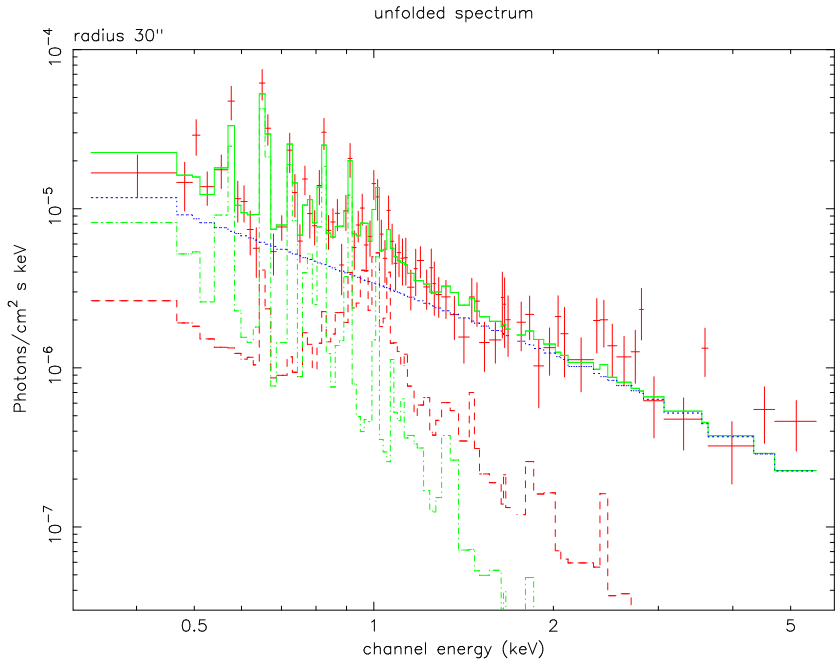}
\caption{Region 30$''$, unfolded spectrum: 0.3 + 1.0 keV APEC + 7 keV
Bremsstrahlung, line-of-sight absorption.}
\label{unfold}
\end{figure}

We find that a single APEC component, with kT$\sim$0.3 keV, and a 7 keV
Bremsstrahlung model are able to approximate the data and give
an acceptable value for the reduced $\chi^2 \sim$ 1.  However, inspection of
the residuals shows a noticeable excess at around  1 keV (with a peak at
about 5 $\sigma$) and a less significant deficit
at 0.6-0.8 keV, as shown in Fig.~\ref{spec1}.
The addition of a second APEC component reduces the $\chi^2$ value and
most of all eliminates the excess, suggesting two separate components
at 0.3 keV and 1.0 keV (Fig.~\ref{spec1}, bottom). Fig.~\ref{contour}
shows the 68\%, 90\% and 99\% confidence regions for the
two interesting parameters, the temperatures of the very soft and soft
components. Fig.~\ref{unfold} shows the best-fit unfolded spectral components,
from which it becomes evident that the $\sim 1$ keV component is clearly dominant over the
$\sim 0.3$ keV component in the
0.7-1.5 keV band.

We tested whether the existing data support a requirement of a second soft plasma component 
in the spectrum, by computing the Bayesian Information Criterion (BIC), discussed by
Schwarz (1978; 
an astronomical introduction to it can be found in Liddle 2004, and an application in Tajer
et al. 2007) for both models.   This quantity approximates the Bayes factor (Jeffreys 1961; Kass
\& Raftery
1995), which gives the posterior odds of one model against another, presuming that the models
are equally favored prior to the data fitting (Liddle 2004). 
The difference $\Delta$BIC can be used to evaluate the relative value of the two models. 
We find $\Delta$BIC of 5-6 (depending on the region used), which 
can be used as a strong evidence (see Jeffreys 1961; Liddle 2004) 
that we are justified in  introducing the 1 keV component in the spectral modelling.

\begin{figure}
\plotone{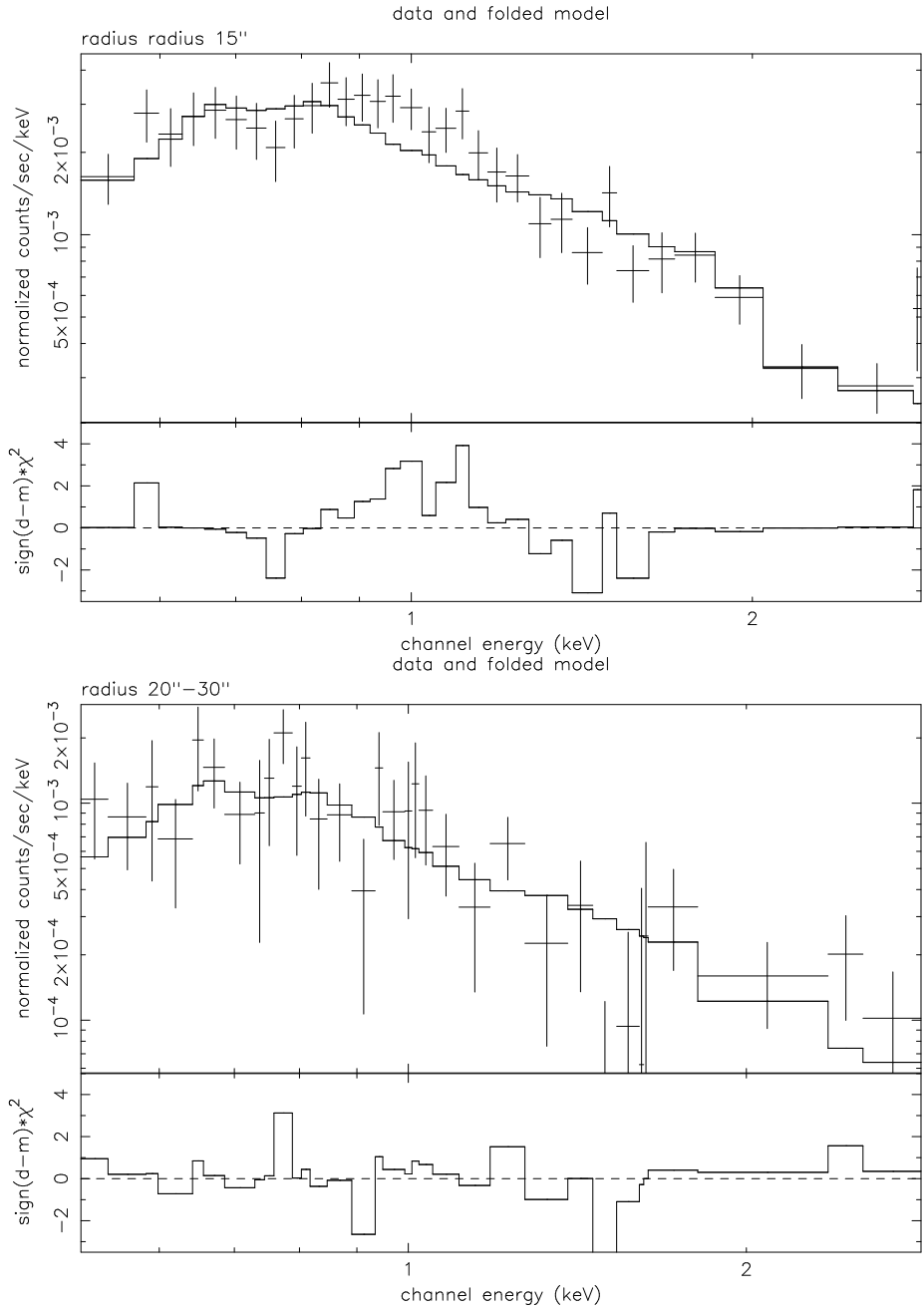}
\caption{Spectral data in the 0\arcsec-15\arcsec\ and
20\arcsec-30\arcsec\ region, fitted with
a single temperature plasma model}
\label{spec2}
\end{figure}

We have investigated further the spatial distribution of the different spectral
components by comparing the  spectra from  the $2''-15''$ and
$20''-30''$ regions, both fitted with a single APEC spectrum at 0.3 keV 
(Fig.~\ref{spec2}). 
It is clear that there is no residual emission at 1 keV in the spectrum
from the outer region; this suggests that the additional ``soft" component
is more concentrated towards the center, consistent with the radial
profile comparison of Fig.~\ref{softprof}.

From the spectral results we can estimate the luminosity of each
individual component, summarized in Table 1. 
The measured unabsorbed
luminosity of the total unresolved emission, within the 30$''$ region,
is $\rm L_x \sim 4 \pm 0.4 \times 10^{38}$ \ergs\ in the 0.3-10 keV range,
equally divided below and above 2 kev.  The hard band luminosity is
totally accounted for by the Bremsstrahlung component, which alone
gives a luminosity L$_x \sim 1\pm 0.1 \times 10^{38}$ \ergs\ in the 0.5-2.0 keV band.
The 1 keV component contributes for  $\rm L_x \sim 4 ^{+2.5}_{-0.2} \times 10^{37}$
\ergs\ in this band, and the residual $\rm L_x \sim 6^{+2.5}_{-0.4}\times 10^{37}$
\ergs\ is due to the 0.3 keV emission.  Errors are estimated from the best fit values (68\% confidence values, from XSPEC).  We have estimated a correction
factor to recover the area lost due to masking the point sources (see
Table 1), and a second one to account for the fraction of emission not
included in the area used in the spectral analysis (given in Table 2).
These can be used to estimate the total luminosity of the very soft
and hard components, which we list in Table 2.  The component at 1 keV
appears to contribute only in the central region, for r $< 30''$.

\section{Summary of results and comparison with the literature}

We have analyzed the spectral and spatial distribution of the  diffuse
X-ray emission in NGC 3379 with deep {\it Chandra} data, cleaned of all
detected sources down to a threshold luminosity L$_x \sim 10^{36}$ \ergs\
(see Brassington et al. 2008).  The residual emission can be detected only
out to  $\rm r \sim 1'$, corresponding to 3 kpc, well within the D$_{25}$,
measured at $5.4'\times4.8'$ (from NED, see also Cappellari et al. 2006).

The spectral distribution of this unresolved emission in the 30\arcsec\
region used for the spectral analysis suggests the presence of 3 distinct
components.  The dominant contribution in the broad 0.3-10 keV band can
be modeled with a Bremsstrahlung emission at $\sim 7 $ keV and  is
likely to arise from lower luminosity LMXBs and other stellar components 
that can not be detected individually in the present data, as also 
suggested by its spatial distribution, which follows
that of the optical and IR stellar light (Fig.~\ref{softprof}). 
This component accounts for all of the emission above 2 keV 
and about 1/2 of the emission below 2 keV, and has an estimated total $\rm L_x
\sim 5 \times 10^{38}$ \ergs\ (0.5-10 keV, see Table 2). 

We also  find two additional optically thin thermal components, with best fit 
temperatures of $\sim 0.3$~keV and $\sim 1$~keV,
contributing roughly 2/3 and 1/3 of the 0.5-2.0 keV residual
luminosity, respectively (see Table 1), in the 30\arcsec\ radius region.
This result  is supported by the combined evidence of  excess emission at
$\sim 1$ keV over
the dominant 0.3 kev plasma in the spectral data (Fig.~\ref{spec1}), 
and of a different spatial distribution of the two components, when
the data below and above 0.7 keV are selected
(Fig.~\ref{softprof}). 

On average, the 0.3 keV component has a spatial distribution consistent
with that of the lower luminosity LMXBs and coronally active stars
(traced by the emission above 1.5 keV) and of the stars (traced by the
I-band and K-band emission), 
while the 1~keV emission is more centrally peaked, 
and is prominent out to $\rm r \sim 15''-20''$. 

Two previous detailed studies of the diffuse emission of NGC~3379 based on  {\it Chandra} data can
be found in the literature. Based on the first 2001 short observation
David et al. (2005)  conclude that the diffuse
emission is dominated by undetected faint
point sources, but with a 10\% contribution from a
gaseous component, with kT$\sim 0.6$ keV,
in the central 770 pc. 
This hot ISM would have a luminosity of $\sim 9
\times 10^{37}$ \ergs (0.3-10 keV) and a mass of $\sim 5
\times 10^{5}$ M$_\sun$. Our estimate of the luminosity of the optically thin thermal
diffuse emission in the same region
is consistent with that of David et al (2005), but we are able to distinguish two
separate components of this emission.
NGC 3379 is also included in the sample examined by Fukazawa et al. (2006), who also report 
a detection of a component at 0.5 keV in a small region of the galaxy. 

More recently, Revnivtsev et al. (2008) have suggested a different
interpretation for the unresolved emission: using the full dataset as
in the present work, they derive a radial profile of the 0.5-2.0 keV
emission, cleaned of detected sources, which they compare with the
K-band profile.  Based on this comparison they conclude that all of
the unresolved emission observed in NGC 3379 can be attributed to the
emission of stellar sources, with no evidence of a truly diffuse ISM.
Our net radial profile of Fig.~\ref{raw} is entirely consistent with that
of Revnivtsev et al. (2008), giving us confidence that slight differences
in the data analysis,  in the subtraction of detected sources obtained
with different algorithms or in the choice of the background level,
are not significant sources of discrepancies.  The main difference
between this work and  Revnivtsev et al. (2008) therefore lies in our
spectral and spatial analysis, that enables us to distinguish two separate
spectral components  of the soft diffuse emission, with different spatial
properties, which contribute equal amounts to the total  in the inner
$15''$ radius.

\section{Discussion}
\label{models}

Our analysis suggests the presence of three components of the diffuse
emission of NGC 3379. As summarized above, the hard and very soft emission
are spatially consistent with a stellar origin; a detailed discussion of
this integrated X-ray emission from stellar sources (LMXBs, cataclysmic
variables, active stars and stellar coronae) is given by Revnivtsev et al
(2008). This stellar emission has been invoked to  explain
the diffuse X-ray emission of the Galactic Ridge (Revnivtsev et al
2006, Sazonov et al. 2006), the bulge of M31 (Bogd\'an \& Gilfanov 2008), M32 (Revnivtsev et
al. 2007) and  NGC821 (Pellegrini et al 2007a). 

In what follows we will concentrate on the emission of the (0.7-1.5)~keV
band.  This component has a different spatial distribution, 
more concentrated in the center and more centrally peaked than the stellar emission (see
Fig.~\ref{softprof}), 
indicating a different origin.  We suggest that
this component is likely to originate from a $\sim$1~keV hot plasma, and discuss
it in the context of an appropriate model for a galactic flow.

\subsection{Evidence for hot ISM in NGC 3379}

\begin{figure}
\plotone{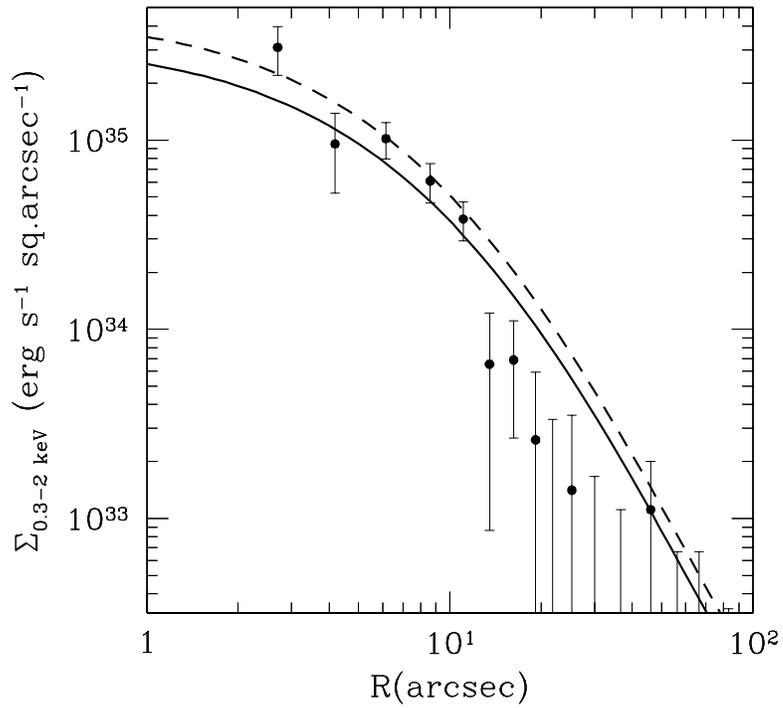}
\caption{The observed 0.3--2 keV emission from the ``pure" 1 keV component, compared with
the profiles resulting for two gas models, one with $s=1.0$ (dashed line)
and the other with $s=1.5$ (solid line).
See Sect.~\ref{models} for more details.
}
\label{gas}
\end{figure}

The interpretation of the 0.7-1.5 keV emission detected in NGC 3379 as
due to a different component than the softer and harder ones opens the
possibility that this is the long sought for evidence of a hot ISM in
this galaxy (e.g., Ciotti et al. 1991, David et al. 2005).
However,  as shown by Fig.~\ref{unfold}, all three components
contribute to the 0.7-1.5 keV emission, while both the softer, and in
particular the harder, energy ranges are less contaminated by the
other components.  The spectral results can be
reliably used to estimate the 
contribution of each component to the total luminosity.  To estimate 
the spatial distribution of each separately, we have to resort to a
few reasonable assumptions, which we have already discussed and that we
summarize below: a) the profiles in both the very soft and hard
energy bands are very similar in shape, therefore we can treat them as
having a common origin with different rescaling factors; 
b) they are also consistent with the  I-band and K- band profiles, again with possibly
different rescaling along the full spatial extent probed; and c)
the  0.7-1.5 keV profile is also consistent in shape with all above
profiles for r$>15''$, where there is no longer any 
evidence of the emission due to the 1 keV component (Fig.~\ref{spec2}). 
We can therefore reasonably conclude that the
0.7-1.5 keV emission at  r$>15''$ traces the soft and hard components, 
and that an extrapolation at smaller radii could be used to give a reasonable estimate of
the local contribution from the stellar component
at different radii. 
We have used the K-band data normalized to the  0.7-1.5 keV emission at r$>15''$ as the best proxy to extrapolate the profile at smaller radii, since we can use it as a model with virtually no errors.  While this relies on the assumption that the stellar light is a good tracer of the X-ray binary population, with no local, though small, variations, the observed agreement with the X-ray profiles of the residual emission at softer and harder bands (Fig.~\ref{softprof}) supports this assumption.  We therefore  derive the radial distribution of the $\sim 1$ keV component
as the excess over our model
(Fig.~\ref{gas}) which we can now compare to the expected emission from
a hot ISM, derived from the hydrodynamical simulations described below.
Note that the X-ray luminosity  
derived from integrating the radial distribution of Fig.~\ref{gas} is L$_x \sim 2.5 \pm 1 \times 10^{37}$ erg s$^{-1}$, consistent with 
that  derived from the spectral analysis.

From the spectral results and the assumption of spherical symmetry, we can estimate the 
average density and total mass of gas responsible for the emission measured. 
We derive a very low density of $\rm n_e \sim 0.007$ cm$^{-3}$ and a total mass of gas M$\sim
3 \times 10^5$ M$_{\sun}$, in a region of $\sim 770$ pc in radius at the center. 

\subsection{The hot gas from hydrodynamical simulations}

The round optical shape (morphological type E1), its optical colors,
typical of an old population (Terlevich \& Forbes 2002), and the lack of
signs of current or past interactions (Schweizer \& Seitzer 1992) allow
us to assume that a simple ``passively evolving" stellar population
feeds a hot gas flow, whose evolution can be reasonably studied with
spherically symmetric hydrodynamical simulations.  The numerical code
used to solve the time-dependent equations of hydrodynamics with source
terms is described in details in Ciotti et al. (1991).  We adopt a
central grid spacing of 20 pc, which provides a good match to the 
scale of the observations in the inner regions;
the simulations do not cover feedback effects and the flow evolution
close to the central massive black hole  (Shapiro et al. 2006).

Given the high quality data available for this galaxy, it is possible to  build 
a mass model tailored 
specifically onto NGC~3379, and consider the best recipes available for
the mass and energy source terms.  These are briefly described below.

\subsection{The mass model}

The optical profile is well described by a de Vaucouleurs (1948) law
over a span of ~10 magnitudes (Capaccioli et al. 1990; Peletier et
al. 1990), therefore the simulations use the stellar mass density profile given
by the Hernquist (1990) distribution, that is a very good
approximation of the de Vaucouleurs law and has the advantage
that its dynamical properties can be expressed analytically.  We also
impose the observed B-band luminosity  of 1.5$\times 10^{10}$ L$\rm _{B\sun}$,
the central stellar velocity
dispersion of 230 km s$^{-1}$ (from the detailed modeling of
ground-based and $HST$ spectroscopy, Shapiro et al. 2006) and a total
stellar mass $M_* = 10^{11}M_{\odot}$,  obtained from
dynamical and stellar population synthesis studies
(Saglia,
Bertin \& Stiavelli 1992, Gerhard et al.  2001, Napolitano et
al. 2005, Cappellari et al. 2006,
Douglas et al. 2007). The resulting shape of the stellar profile, when projected,
is close to the optical profiles in Fig.~\ref{softprof}. 

The radial density distribution of the dark haloes of ellipticals is
not well constrained by observations; theoretical arguments  and
high resolution numerical simulations of dissipation-less collapse
produce a density distribution $\propto r^{-1}$ near the center
(Dubinsky \& Carlberg 1991, Navarro, Frenk, \& White 1996).  The model
galaxy is therefore a superposition of two Hernquist density
distributions, one for the luminous matter and one for the dark matter.

The total amount of mass in NGC~3379 has been calculated several times
using the observed stellar velocity dispersion profile,
extending out to $\sim 1-2$ effective radii (\Reff, e.g., Saglia, Bertin \& Stiavelli
1992, Kronawitter et al. 2000, Samurovic \& Danziger 2005), and dynamical
tracers as planetary nebulae or globular clusters extending out to
larger radii (Romanowsky et al. 2003, Dekel et al. 2005, Teodorescu et al. 2005; Pierce
et al. 2006, Bergond et al. 2006, Douglas et al. 2007, De Lorenzi et al. 2008).  Regardless of the
details of the modelling, studies based on the observed
stellar velocity dispersion profile invariably show very little dark
matter within \Reff, or even no dark matter.  
Within 2--3   \Reff,
the stellar component alone satisfies reasonably well the
observations, and the amount of dark matter is constrained to be at
most very modest (with a dark to stellar mass ratio of the order of $M_h/M_*\sim 0.1$, 
Kronawitter et al. 2000; Cappellari et al. 2006). The evidence for dark matter is at
the largest scales (10--40 kpc), 
probed mostly by globular clusters (see also
the analysis of an HI ring, Schneider 1985), with a global galactic
value of $M_h/M_*=1.1-1.5$ (Dekel et al. 2005), $M_h/M_*\sim 2$ (Puzia
et al. 2004), up  to
$M_h/M_*\sim 6$ (Bergond et al. 2006).  

In order to make the model mass distribution consistent with the
observations, we also imposed the ratio $M_h/M_* = 0.1$ within one 
\Reff\footnote{In the code the presence of dark matter
contributes also to produce the observed central stellar velocity
dispersion, as described in Pellegrini \& Ciotti (2006), though this
aspect is not very important here given the small amount of dark
matter in the central regions.}, while on the global galactic scale
it has been varied instead between 1 and 6.

\subsection{Time evolving inputs}

The model does not assume a steady state configuration, therefore the  input ingredients 
of the numerical simulations, which are
the rates of stellar mass loss from the aging stellar population and
the rate of SNIa heating, are both evolving with time.  In the numerical code the exact mass return
rate prescribed by the stellar evolution theory is used 
(Ciotti et al. 1991), updated to
take into account the latest stellar population synthesis models
and more recent estimates for the mass ejection from
stars as a function of their mass (Maraston 2005).
This gives a present epoch ($\sim 10$ Gyr) mass return rate of
$\sim 0.3 M_{\odot}$ yr$^{-1}$ for the whole galaxy.

The SNIa heating rate is parameterized as $L_{SN}(t)= E_{SN}\,
R_{SN}(t)\,L_B$, where $E_{SN}$ is the kinetic energy injected in the
ISM by one SNIa, and the number of events as a function of time is
$R_{SN}(t)L_B\propto t^{-s}$, where the slope $s$ describes the
unknown decay rate.  $R_{SN}(t)$
is normalized to give the present SNIa's explosion rate in nearby
E/S0s of Cappellaro et al. (1999), i.e., 0.16$h_{70}^2$ SNu (where 1 SNu = 1 SNIa per 100 yrs per
$10^{10}L_{B,\odot}$, $h_{70}=H_{\circ}/70$), that has an associated
uncertainty of $\sim 30$\%.  A detailed
theoretical modeling of the evolution of the rate shows that, after
the first 0.5--1 Gyr, it is well approximated by a power law with a
slope $s\sim 1.0-1.2$ for double degenerates exploders, and a slope
$s\sim 1.5-1.7$ for the single degenerate exploders (Greggio 2005).

The thermalization of the stellar mass losses to the local ``temperature"
set by the stellar velocity dispersion is another heating source, though of lower 
relevance.  Its contribution is determined  at each radius 
using the velocity dispersion profile obtained by solving
the Jeans equation for the two-component Hernquist model in the
globally isotropic case (Pellegrini \& Ciotti 2006).  For realistic
anisotropy distributions, the difference with the isotropic case is small; 
in addition, the heating due to thermalization of stellar motions is significantly 
smaller than SNIa heating.

\subsection{Results and comparison with observations}

With the SNIa's explosion rate from Cappellaro et al. (1999) and the
adopted mass model, the resulting gas flow phase is of a global
outflow, with velocities ranging from $10^7$ to a few $10^7$ cm s$^{-1}$,
going from the central region  to the galaxy outskirts (e.g., subsonic out to r$\sim$1 kpc and
supersonic outwards).  When varying the total
amount of dark matter within the range $M_h/M_*=1-6$ allowed for by
observations, this flow pattern keeps basically the same (provided that the SNIa's rate increases together with $M_h/M_*$,
but still within its observational uncertainty).

For the resulting gas flows, the X-ray luminosity in the energy bands
of 0.3--2 keV and 2--8 keV, and the temperature weighted with the
emission in the 0.3--8 keV band ($<kT>$), have been calculated.  In order to
do this, the
values of the cooling function $\Lambda(T,Z)$, as a function of
temperature, have been obtained with the APEC code in XSPEC, for solar
abundance, consistent with the estimate for the stellar population
(Terlevich \& Forbes 2002); a ``best fit" $\Lambda$ was then  derived 
from these values and used in
the $L_X$ and $<kT>$ calculation (that takes into account the whole
density and temperature distributions over the computational grid).

After $\sim 9$ Gyr, the gas has a 
0.3--2 keV luminosity of $2\times
10^{37}$ erg s$^{-1}$, in good agreement with the observed one.  This
is a reasonable time scale that corresponds to the estimated
age of NGC 3379:  Terlevich \& Forbes (2002) give an age of
$9.0^{+2.3}_{-1.9}$ Gyr; more recently, a mean age in the range 8--15
Gyr has been confirmed by Gregg et al. (2004) and Denicol\`o et
al. (2005).  The 
central temperature is $\sim 10^7$ K out to a few hundred pc, and
is steeply declining outwards; the central gas density is
$n_e=2\times 10^{-2}$ cm$^{-3}$ and is also steeply declining
outwards. The mass of the hot gas is $\sim 3-4\times 10^5 M_{\odot}$ within
a radius of 800 pc (in good agreement with that obtained in Sect. 4.1). The
emission weighted gas temperature within a radius of $15^{\prime\prime}$ is $<kT>\sim 0.8$ keV, a value near the  lower range of those allowed for by observations (see Fig.~\ref{contour}).  The
projection along the line of sight of the 0.3--2 keV emission is shown
in Fig.~\ref{gas}, for two models with a total $M_h/M_*=4$, a SNIa's
rate close to that of Cappellaro et al. (1999), and a slope for the SNIa's decay rate of $s=1.0$ and $s=1.5$.  Both model profiles reproduce well
the observed ``gas profile" derived above. We note that a larger age for the gas flow
would result in a lower stellar mass loss rate and lower gas content,
with consequently lower ISM emission, below the observed one.

\section{Conclusions}

With the very deep {\it Chandra} observation of NGC 3379 we have 
scored two firsts:  we have the first evidence of a very small amount of 
hot ISM in a nearby galaxy, down to the level of $\rm L_x \sim 4 \times 10^{37}$ \ergs\
(0.5-2.0 keV), and at the same time we have the first evidence that this 
gas is in an outflow phase, which has been predicted by models but as of yet
has not been observed.
As a third and equally important result, we find that detailed modelling based on the
optical data (stellar mass, total mass and supernova rate) reproduces remarkably
well the observed total
luminosity, total mass  and spatial distribution of the hot phase of the gas. 

After detection and removal of point sources down to a luminosity of $\rm L_x \sim 10^{36}$
\ergs\ we have been able to study the unresolved emission in NGC 3379. 
The detailed spectral analysis has led us to the evidence of three separate components: 
a very soft (plasma at kT$\sim$0.3 keV) and a hard (Bremsstrahlung at kT=7 keV)
components can be attributed to the integrated emission of 
unresolved sources, closely linked to the stellar population (binary systems and coronally
active stars, as discussed by Revnivtsev et al. 2008).  However we also  detect
a hot ISM in this galaxy, within the innermost $\sim$ 800 pc, with a
luminosity $\rm L_x \sim 4 \pm 1.5 \times 10^{37}$ \ergs\ (0.5-2.0 keV). This is 
lowest detection of a hot gas in a nearby early-type galaxy.  The  total
gas mass involved is also very small, of a few $10^5$ M$_{\sun}$.

With the aid of hydrodynamical simulations and a galaxy model tailored specifically on 
NGC3379, we have reproduced both  the luminosity and
the average radial distribution of this hot gas component.
For a SNIa's explosion rate consistent with current estimates, the gas
is outflowing,  even for the
"maximum" dark matter model, where the amount of dark mass is the
maximum allowed for by optical studies. 


\section{Appendix: Consideration upon the effects of the PSF}.

At the suggestion of the referee, we have thoroughly checked that the results obtained above, in particular the detection of the $\sim 1$ keV component, are not due to a faulty subtraction of the detected sources. We have run several tests, which we illustrate below.

\noindent 1) We have first redone the spatial and spectral analysis assuming  a circle of $2.5"$ and $3" $ radius for point source exclusion, with analogous results, albeith somewhat larger errors,
due to the smaller statistics available.

\noindent 2) We have also simulated the {\it Chandra}'s  PSF using ChaRT (Carter et al. 2003).  Since we were interested in the wings of the PSF, we have run several simulations for 20,000 counts each and then merged the resulting images to increase the statistics at radii larger than $1''$   We have used the average spectrum obtained in the inner 15$''$, inclusive of all detected  sources,  as an input parameter in all simulations.  We have then treated the results as we have the data of NGC 3379, namely we have produced radial profiles in the same energy bands and calculated the expected contribution from point sources in the regions of interest.  We expect that 4\% of the counts detected from a point sources within 2$''$ from its  peak falls in the $2''-30''$ region, and 3.6\% in the $2''-15''$  region, in the  0.3-2.0 keV band (with values between 3\% and 5\% in the softer and harder of the bands considered).   If we integrate the contribution due to the ``wings'' of  all point sources detected in a radius=30\arcsec,  we can only account for 10-25\% of the detected counts in the 0.3-0.7, and 0.7-1.5 keV bands in the unresolved component.  This is a generous estimate of the contribution from the discrete population, since only sources at the center of the region will contribute the full amount, the wings of the others being only partially included in the region considered.

\noindent 3)  We have used the current documents online\footnote{{\tt http://cxc.harvard.edu/cal/Hrma/psf/index.html} } and directly contacted the experts on both the PSF and its wings (Dr. T Gaetz and D. Jerius) to better understand the effects of the wings of the point spread function.   The most updated studies on the PSF wings are based on the ground XRCF and on-orbit Her X-1 data. Both data sets demonstrate the existence of the PSF wings, but the effects of this component decrease rapidly with decreasing energy. The ray trace model used to simulate the PSF (point 2 above) includes some of the effects of wings, and appears to be reliable  at low energies ($<$ 2 keV) and small radii, although it might be underestimating the wing effects at energies above 2.0-2.5 keV.  However, we  are mostly interested in the emission below $\sim$ 1.5 keV and relatively small radii, where the model appears to be adequate.

\noindent 4) We have re-analyzed the spectrum of the emission in the $2''-15''$ regions excluding only the 5 brightest sources with a significantly smaller exclusion region of $1''$ radius.  This exclusion is necessary to avoid that the point source population completely dominates the spectral results; on the other hand, if the sources  were responsible for the 1 keV component, it should be significantly stronger when sources are included in the spectral analysis. 
The spectral results indicate that, while the net counts increase by a factor of 4, the 1 keV component only increases by a factor of 1.5 the original value, consistent with the ratio of the areas of the regions considered.  
 
All in all, we believe that the excess emission observed in the 0.7-1.5 keV band is a real effect, and is not an artifact related to the wings of the many individual sources detected in the area.  This excess is relative to the emission in adjacent energy bands, and is seen at energies where the effects of PSF wings are significantly smaller than at higher energies.  We also believe that there is excess emission over the scattering from the  detected point sources, although we are fully aware of the fact that the exact evaluation of its strength has large uncertainties, both in the soft and in particular in the hard band.

\acknowledgments
We thank Stefano Andreon for many interesting discussion on the statistical 
aspects of the data analysis, and Terry Gaetz and Diab Jerius for their help in understanding the PSF issues. 
We thank the referee for raising the issue of the effects of the PSF, that prompted us to better investigate the reliability of the results.  
The data analysis was supported by the CXC CIAO software and CALDB, and
has made use of the SAOImage DS9 and funtools softwares, developed by 
the Smithsonian Astrophysical Observatory. We
 have used the NASA NED and ADS facilities, and have extracted archival
 data from the {\it Chandra}, HST and 2MASS archives. 
 This work was supported by the {\it
 Chandra} GO grant G06-7079A (PI: Fabbiano) and sub-contract G06-7079B
 (PI: Kalogera). We acknowledge partial support from NASA contract
 NAS8-39073 (CXC); A. Zezas acknowledges support from NASA LTSA grant
 NAG5-13056; S. Pellegrini acknowledges partial financial support from
 the Italian Space Agency ASI  (Agenzia Spaziale Italiana) through grant
 ASI-INAF I/023/05/0.



{\it Facilities:}  \facility{CXO (ACIS)}.

\clearpage




\clearpage

\begin{table}
\begin{center}
\caption{Measured luminosities of the different components}
\begin{tabular}{ccccccc}
\tableline\tableline
Region & Net counts\tablenotemark{a} & L$_X$\tablenotemark{b} & L$_X$\tablenotemark{b} &
L$_X$\tablenotemark{b} & B/A1/A2\tablenotemark{c} & Correction\tablenotemark{d} \\
&&0.3-10 keV & 0.5-2.0 keV & 2.0-10.0 keV& 0.5-2.0 keV \\
\tableline
2\arcsec-30\arcsec & 1550$\pm$53 & 4.4$\times10^{38}$ &2.0$\times10^{38}$ & 2.0$\times10^{38}$  &
1/0.6/0.35 &  1.15\\
2\arcsec-15\arcsec &978$\pm$35 	&2.9$\times10^{38}$ &1.3$\times10^{38}$
&1.3$\times10^{38}$ &0.8/0.27/0.28 &1.34\\
\tableline
\end{tabular}
\tablenotetext{a}{Net counts used in the spectral fit}
\tablenotetext{b}{Total luminosity in the given band in \ergs}
\tablenotetext{c}{$\rm L_x$  in the 0.5-2  keV band for the Bremsstrahlung / APEC @ 0.3 keV/ APEC @ 1 keV components, in units of $10^{38}$ \ergs}
\tablenotetext{d}{Correction factor to take into account the area lost due to the excised
sources}
\end{center}
\end{table}

\begin{table}
\caption{Total luminosities of the different components, after correcting
for area lost and total extent 
}
\begin{tabular}{lcccccc}
\\
\tableline\tableline
Component & Extension & Correction\tablenotemark{a}& L$_X$ & L$_X$ \\ 
          & (arcsec) & & 0.5-2.0 & 2-10 \\
\tableline
Very soft (0.3 keV)&  95 & 1.60& 1.1$\times10^{38}$& 2.3$\times10^{35}$\\
Soft (1 keV)    & 20 & 1.00 & 4.0$\times10^{37}$ & 5.0$\times10^{36}$  \\
Hard (7 keV)    & 50 & 1.35 & 1.6$\times10^{38}$& 3.0$\times10^{38}$\\
\tableline
\end{tabular}
\tablenotetext{a}{Correction factor to extrapolate from the 
spectral extraction region (30\arcsec, see Tab. 1) to the total source
counts}
\end{table}


\begin{thebibliography}{}

\bibitem{} Bergond, G., Zepf,
S.~E., Romanowsky, A.~J., Sharples, R.~M., \& Rhode, K.~L.\ 2006, \aap,
448, 155

\bibitem{} Brassington, N. J. et al 2008, ApJ, submitted.

\bibitem{} Bogdan, A., \& Gilfanov, M.\ 2008, ArXiv e-prints, 803, arXiv:0803.0063 

\bibitem{} Canizares, C.~R., 
Fabbiano, G., \& Trinchieri, G.\ 1987, \apj, 312, 503 

\bibitem{}Capaccioli, M., Held, E.V., Lorenz, H., Vietri, M. 1990, AJ 
99, 1813

\bibitem{}Cappellari, M., Bacon, R., Bureau, M., et al. 2006, MNRAS 366, 
1126

\bibitem{}Cappellaro, E., Evans, R., Turatto, M. 1999, A\&A 351, 459

\bibitem{}Carter, C. Karovska, 
M., Jerius, D., Glotfelty, K., 
\& Beikman, S.\ 2003, Astronomical Data Analysis Software and Systems XII, 295, 477 

\bibitem{} Ciotti, L., D'Ercole, 
A., Pellegrini, S., \& Renzini, A.\ 1991, \apj, 376, 380

\bibitem{} David, L.~P., Forman, W., 
\& Jones, C.\ 1991, \apj, 380, 39

\bibitem{}  David, L.~P., Jones, C., 
Forman, W., \& Murray, S.~S.\ 2005, \apj, 635, 1053 

\bibitem{}Dekel, A., Stoehr, F., Mamon, G. A., Cox, T. J., Novak, G. S.,
Primack, J. R. 2005, Nature 437, 707

\bibitem{}De Lorenzi, F.; Gerhard, O.; Coccato, L. et al. 2008, 
arXiv:0804.3350

\bibitem{}de Vaucouleurs G., 1948, Ann.Ap., 11, 247

\bibitem{} Denicol{\'o}, G., 
Terlevich, R., Terlevich, E., Forbes, D.~A., 
\& Terlevich, A.\ 2005, \mnras, 358, 813 

\bibitem{} Douglas, N.~G., et al.\ 
2007, \apj, 664, 257

\bibitem{}Dubinski, J., Carlberg, R. G. 1991, ApJ 378, 496

\bibitem{} Fabbiano, G., et al.\ 
2006, \apj, 650, 879 

\bibitem{} 2006, \araa, 44, 323 

\bibitem{} Fabian, A.C., \& Canizares, C.R. 1988, Nature, 333, 829

\bibitem{} Forman, W., Jones, C., 
\& Tucker, W.\ 1985, \apj, 293, 102

\bibitem{} Fukazawa, Y., 
Botoya-Nonesa, J.~G., Pu, J., Ohto, A., \& Kawano, N.\ 2006, \apj, 636, 698 

\bibitem{}Gerhard O., Kronawitter, A., Saglia, R. P., Bender, R. 2001, 
AJ 121, 1936

\bibitem{} Gregg, M.~D., Ferguson, 
H.~C., Minniti, D., Tanvir, N., \& Catchpole, R.\ 2004, \aj, 127, 1441

\bibitem{}Greggio, L., 2005, A\&A 441, 1055

\bibitem{}Hernquist L.E., 1990, ApJ, 536, 359

\bibitem{}Ho, L.C. 2008, arXiv:0803.2268

\bibitem{}Jeffreys H., 1961, Theory of probability, 3rd edn, Oxford Univ. Press, Oxford

 \bibitem{}Kass R. E., Raftery A. E. , 1995, Journ. American Stat. Assoc., 90, 773

\bibitem{}     Kim, D.-W., \& Fabbiano, G.\ 2003, \apj, 586, 826 

\bibitem{} Kim, D.-W., et al.\ 2006, 
\apj, 652, 1090 


\bibitem{} Kim, D.-W., Fabbiano, G., 
\& Trinchieri, G.\ 1992, \apj, 393, 134 

\bibitem{}Kronawitter A., Saglia R.P., Gerhard O., Bender R., 2000,
          A\&AS, 144, 53

\bibitem{} Liddle, A.~R.\ 2004, \mnras, 
351, L49 

\bibitem{}Maraston, C. 2005, MNRAS 362, 799

\bibitem{} Matsushita, K., et 
al.\ 1994, \apjl, 436, L41 

\bibitem{} Napolitano, N.~R., 
et al.\ 2005, \mnras, 357, 691

\bibitem{}Navarro, J.F., Frenk, C.S., White, S.D. M. 1996, ApJ 462, 563

\bibitem{}Peletier R.F., Davies, R.L., Illingworth, G., Davis, L.E., 
Cawson, M. 1990, AJ 100, 1091

\bibitem{} Pellegrini, S.\ 2005, \apj, 
624, 155

\bibitem{}Pellegrini, S., Ciotti, L. 1998, A\&A, 333, 433


\bibitem{}Pellegrini, S., Ciotti, L., 2006, MNRAS 370, 1797

\bibitem{}  Pellegrini, S., 
Siemiginowska, A., Fabbiano, G., Elvis, M., Greenhill, L., Soria, R., 
Baldi, A., \& Kim, D.~W.\ 2007a, \apj, 667, 749 

\bibitem{} Pellegrini, S., 
Baldi, A., Kim, D.~W., Fabbiano, G., Soria, R., Siemiginowska, A., 
\& Elvis, M.\ 2007b, \apj, 667, 731 

\bibitem{} Pellegrini, S., \& Fabbiano, G.\ 1994, \apj, 429, 105 

\bibitem{}Pierce, M., Beasley, M. A., Forbes, D. A., et al. 2006, MNRAS 366,
1253

\bibitem{} Puzia, T.~H., et al.\ 2004, \aap, 415, 123

\bibitem{} Revnivtsev, M.,
Sazonov, S., Gilfanov, M., Churazov, E., \& Sunyaev, R.\ 2006, \aap,
452, 169

\bibitem{} Revnivtsev, M., Churazov, E.,
Sazonov, S., Forman, W., \& Jones, C.\ 2007, \aap, 473, 783

\bibitem{}Revnivtsev, M., Churazov, E.,
Sazonov, S., Forman, W., \& Jones, C.\ 2008,  ArXiv e-prints, 804, arXiv:0804.0319

\bibitem{}Richstone, D.; Ajhar, E. A.; Bender, R.; Bower, G.; Dressler, 
A.; Faber, S. M.; Filippenko, A. V.; Gebhardt, K.; Green, R.; Ho, L. C.; 
Kormendy, J.; Lauer, T. R.; Magorrian, J.; Tremaine, S.
1998 Nature 395 14

\bibitem{}Romanowsky, A. J., Douglas, N. G., Arnaboldi, M., et al. 2003,
Science 301, 1696

\bibitem{}Saglia R.P., Bertin G., Stiavelli M., 1992, ApJ, 384, 433

\bibitem{}Samurovic, S.; Danziger, I. J. 2005, MNRAS 363, 769

\bibitem{} Sazonov, S., Revnivtsev, M., Gilfanov, M., Churazov, E., \& Sunyaev, R.\ 2006, \aap, 450, 117

\bibitem{} Schneider, S.\ 1985, \apjl, 
288, L33

\bibitem{}Schweizer, F., Seitzer, P. 1992, AJ 104, 1039

\bibitem{}Schwarz G., 1978, Annals of Statistics, 5, 461

\bibitem{} Shapiro, K.~L., 
Cappellari, M., de Zeeuw, T., McDermid, R.~M., Gebhardt, K., van den Bosch, 
R.~C.~E., \& Statler, T.~S.\ 2006, \mnras, 370, 559

\bibitem{}Springel, V., Di Matteo, T., Hernquist, L. 2005, ApJ 620, L79

\bibitem{} Tajer, M., et al.\ 2007, \aap, 467, 73

\bibitem{}Teodorescu, A. M., Mendez, R. H., Saglia, R. P., Riffeser, A.,
Kudritzki, R.-P., Gerhard, O. E., Kleyna, J. 2005, ApJ 635, 290

\bibitem{}Terlevich, A. I., Forbes, D.A. 2002, MNRAS 330, 547

\bibitem{}Tonry, J.L., Dressler, A., Blakeslee, J.P., et al. 2001, ApJ 546,
681

\bibitem{} Trinchieri, G., \& Fabbiano, G.\ 1985, \apj, 296, 447 


\bibitem{} Wilms, J., Allen, A., 
\& McCray, R.\ 2000, \apj, 542, 914 



\end{thebibliography}
\end{document}